\documentclass[12pt]{article}
 \usepackage{epsfig}
 \def\be{\begin{equation}}
 \def\ee{\end{equation}}
 \def\bea{\begin{eqnarray}}
 \def\eea{\end{eqnarray}}
 \usepackage{graphicx}% Include figure files

 \catcode`\@=11
 \def\lsim{\mathrel{\mathpalette\@versim<}}
 \def\gsim{\mathrel{\mathpalette\@versim>}}
 \def\@versim#1#2{\vcenter{\offinterlineskip
 \ialign{$\m@th#1\hfil##\hfil$\crcr#2\crcr\sim\crcr } }}
 \catcode`\@=12

 \catcode`@=12
\begin{document}
 \thispagestyle{empty}
 \begin{flushright}
 UCRHEP-T632\\
 June 2026\
 \end{flushright}
 \vspace{0.6in}
 \begin{center}
 {\Large \bf Flavor Alignment and Mass Hierarchy: \\ 
Doing Everything Scotogenically \\}
 \vspace{1.5in}
 {\bf Ernest Ma\\}
 \vspace{0.1in}
{\sl Department of Physics and Astronomy,\\ 
University of California, Riverside, California 92521, USA\\}
 \vspace{1.2in}

%{\it (in memory of Eileen)\\}
\end{center}

\begin{abstract}\
Flavor alignment and mass hierarchy in quarks are shown to be achievable 
together in a renormalizable theory using the dark sector, while keeping 
only the one Higgs doublet of the standard model.
\end{abstract}

\vspace{1.5in}

\newpage

\baselineskip 24pt
\noindent \underline{\it Introduction}~:~ 
Structure of the three families of quarks has long been a puzzle in particle 
physics.  There are two issues which are hard to explain.  In the standard 
$SU(3)_C \times SU(2)_L \times U(1)_Y$ gauge model with one scalar Higgs 
doublet (SM), the two $3 \times 3$ quark mass matrices ${\cal M}_u$ and 
${\cal M}_d$ are arbitrary to begin with.  Each is diagonalized by two 
independent unitary transformations:
\begin{equation}
{\cal M}_u = U^\dagger_L \pmatrix{m_u & 0 & 0 \cr 0 & m_c & 0 \cr 0 & 0 & m_t} 
U_R, ~~~ 
{\cal M}_d = D^\dagger_L \pmatrix{m_d & 0 & 0 \cr 0 & m_s & 0 \cr 0 & 0 & m_b} 
D_R. 
\end{equation}
The observed charged-current mixing matrix is then
\begin{equation}
V_{CKM} = U^\dagger_L D_L.
\end{equation}
Since $U_L$ and $D_L$ are not constrained by any symmetry of the SM, why 
should $V_{CKM}$ be very close to being the identity matrix?  Furthermore, 
why should the masses be arranged so that the first family has the smallest 
masses, the second family has larger masses, and the third family the largest 
of them all?  These well-known problems of flavor alignment and mass 
hierarchy have been under study for a long time.  The former usually calls 
for a non-Abelian (discrete) family symmetry, and the latter is often 
interpreted as coming from nonrenormalizable higher-dimensional operators. 
As such, the two approaches appear to be mutually exclusive.

Here it will be shown how everything may actually fit together 
within a \underline{renormalizable} framework, using the scotogenic 
mechanism~\cite{m06,t96} and just the one SM Higgs doublet~\cite{m25} which 
breaks $SU(2)_L \times U(1)_Y$ to $U(1)_Q$.  As a reference, the SM particle 
content is listed in Table~1.

\begin{table}[tbh]
\centering
\begin{tabular}{|c|c|c|c|}
\hline
particle & $SU(3)$ & $SU(2)$ & $U(1)$ \\
\hline
$(u,d)_L$ & 3 & 2 & 1/6 \\ 
$u_R,d_R$ & 3 & 1 & $2/3,-1/3$ \\ 
$(\nu,e)_L$ & 1 & 2 & $-1/2$ \\ 
$e_R$ & 1 & 1 & $-1$ \\ 
\hline
$(\phi^+,\phi^0)$ & 1 & 2 & 1/2 \\ 
\hline
\end{tabular}
\caption{Standard model particles.}
\end{table}

\noindent \underline{\it Dark Sector}~:~ 
The existence of dark matter in the Universe is well-known, but its nature 
remains hidden.  Here a $Z_2$ dark sector is postulated~\cite{m25-3} with heavy 
vectorlike quarks of charges 2/3 and $-1/3$ as the known quarks:
\begin{equation}
\pmatrix{a \cr v}_{L,R} \sim (3,2,1/6), ~~~ a'_{L,R} \sim (3,1,2/3), ~~~ 
v'_{L,R} \sim (3,1,-1/3).
\end{equation} 
Their purpose is to allow one-loop radiative masses for the first and second 
families of quarks, with the help of neutral scalar dark flavons as shown 
in Fig.~1.
\begin{figure}[htb]
\vspace* {-3.5cm}
\hspace*{-3cm}
\includegraphics[scale=1.0]{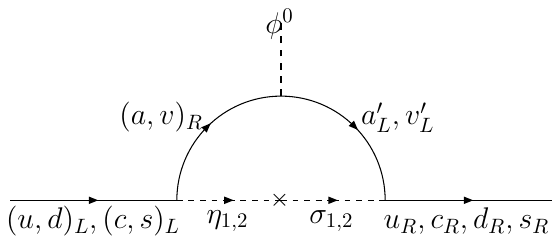}
\vspace{-22.5cm}
\caption{One-loop scotogenic quark masses.}
\end{figure}
Note that it differs from a previously proposed framework~\cite{m14} by 
switching the dark bosons and fermions in the loop. Details will be discussed 
in the following sections.

\noindent \underline{\it Flavor Alignment}~:~ 
Noting that flavor alignment is only needed for left-handed quarks, the 
ovbvious approach is to assume that they transform under a symmetry which 
has at least a doublet representation.  As an example, consider the 
non-Abelian discrete symmetry $S_3$ and let
\begin{eqnarray}
[(u,d)_L,(c,s)_L] \sim 2, ~~~ (t,b)_L \sim 1, \\ 
u_R, d_R, c_R, s_R \sim 1', ~~~ t_R, b_R \sim 1, 
\end{eqnarray}
then only $t$ and $b$ acquire tree-level masses $m_t$ and $m_b$ through the 
one Higgs doublet, whereas the other four quarks remain massless. The 
scotogenic mechanism of Fig.~1 now comes into play. The key is to have
\begin{equation}
(\eta_1,\eta_2) \sim 2, ~~~ \sigma_{1,2} \sim 1'
\end{equation}
under $S_3$, which is softly broken by explicit $\eta-\sigma$ mixing. 
The invariant coupling $(\bar{a}_R u_L + \bar{v}_R d_L) \eta_1^* + 
(\bar{a}_R c_L + \bar{v}_R s_L) \eta_2^*$ then ensures the alignment of 
$(u,c)_L$ with $(d,s)_L$ in $V_{CKM}$ as desired.

\noindent \underline{\it Mass Hierarchy}~:~ 
Whereas the radiative quark masses are expected to be suppressed relative 
to the tree-level masses, the mass hierarchy between the first and second 
families is yet to be explained. Suppose $u_R,d_R,\sigma_1$ have an additional 
imposed global $U(1)_{FN}$ charge~\cite{fn79}, then $\eta_2-\sigma_2$ 
mixing breaks only $S_3$, whereas $\eta_1-\sigma_1$ mixing also breaks 
$U(1)_{FN}$.  The mass hierarchy $m_{u,d} < m_{c,s} < m_{t,b}$ is now explained. 
As to the differences between masses within each family, they come from 
the variation of the Yukawa couplings such as $\bar{u}_R a'_L \sigma_1$ and 
$\bar{d}_R v'_L \sigma_1$ within an order of magnitude.

\noindent \underline{\it Quark Family Structure}~:~ 
With only $\eta_1-\sigma_1$ and $\eta_2-\sigma_2$ mixing, both ${\cal M}_u$ 
and ${\cal M}_d$ are diagonal and $V_{CKM}$ is the $3 \times 3$ identity 
matrix.  This pattern is protected by three separate conserved $U(1)$ 
family symmetries, so the deviations are expected to be naturally small. 
Furthermore, the fact that $u$ goes with $d$, and not $s$ or $b$, is the 
result of the scotogenic framework together with $U(1)_{FN}$.  To obtain 
the observed $U_{CKM}$, ${\cal M}_u$ and ${\cal M}_d$ should have 
off-diagonal entries. Allowing all four dark flavons ($\eta_{1,2}$ and 
$\sigma_{1,2}$) to mix only works for the first and second families. 
To mix them with the third family, one additional dark flavon $\sigma_3 \sim 1$ 
is needed. It couples to $t$ and $b$ through $a$ and $v$, and links them with 
the other quarks in one loop through its mixing with the other flavons.

\noindent \underline{\it Anomalous Higgs Couplings}~:~ 
As first pointed out many years ago~\cite{fm14}, if some entries of the quark 
mass matrices are radiative in origin, the Higgs coupling matrices will not 
be proportional to the mass matrices.  In particular, there should be 
off-diagonal Higgs couplings which are absent in the SM. This offers a 
possible experimental test of the present proposal.

As an illustration, suppose the $2 \times 2$ mass matrix linking 
$(\bar{c}, \bar{t})_L$ to $(c,t)_R$ is given by
\begin{equation}
{\cal M}_{ct} = \pmatrix{ c_L & s_L \cr -s_L & c_L} \pmatrix{m_c & 0 \cr 0 & m_t} 
= \pmatrix{c_L m_c & s_L m_t \cr -s_L m_c & c_L m_t},
\end{equation}
then the Higgs coupling matrix may be written as
\begin{equation}
{\cal Y}_{ct} = {1 \over v\sqrt{2}} \pmatrix{ (1+\epsilon_1)c_L m_c & 
(1+\epsilon_2)s_L m_t \cr -(1+\epsilon_3)s_L m_c & c_L m_t},
\end{equation}
where $\epsilon_{1,2,3}$ come from contributions other than Fig.~1 as 
pointed out in Ref.~\cite{fm14}.  In the mass diagonal basis, 
\begin{equation}
\pmatrix{ c_L & -s_L \cr s_L & c_L} {\cal Y}_{ct} = {1 \over v\sqrt{2}} 
\pmatrix{ (1+\epsilon_1 c_L^2 + \epsilon_3 s_L^2)m_c & \epsilon_2 s_L c_L m_t \cr 
(\epsilon_1-\epsilon_3)s_L c_L m_c & (1+\epsilon_2 s_L^2) m_t}.
\end{equation}
This means that the rare decay $t \to hc$, which proceeds in the SM only 
through a loop and is very much suppressed, is enhanced a great deal and 
could be experimentally observed.

The most recent limit~\cite{CMS25} on the branching fraction of $t \to hc$ 
is $4.3 \times 10^{-4}$.  Using the analysis of Ref.~\cite{gjk21}, this 
translates to the bound
\begin{equation}
\epsilon_2 s_L c_L < 0.057.
\end{equation}
The mixing parameter $s_L$ is unknown but probably of order 
$|V_{cb}| \sim 0.04$ and $\epsilon_2$ may be of order 1.   Hence this 
decay is potentially observable in the future as the intensity of the 
Large Hadron Collider improves.

\noindent \underline{\it Concluding Remarks}~:~ 
The seemingly opposing phenomena of flavor alignment and mass hierarchy in the 
quark sector are resolved with the help of the scotogenic mechanism.  The 
resulting framework has dark matter candidates and just the one Higgs boson. 
It is nevertheless not the same as in the standard model, because it has 
anomalous couplings and in particular, the $t \to hc$ decay is likely to 
be observable.

\newpage
\bibliographystyle{unsrt}

\begin{thebibliography}{99}
\bibitem{m06} E. Ma, Phys. Rev. {\bf D73}, 077301 (2006).
\bibitem{t96} Z. Tao, Phys. Rev. {\bf D54}, 5693 (1996).
\bibitem{m25} E. Ma, Mod. Phys. Lett. {\bf A40}, 2550200 (2025).
\bibitem{m25-3} E. Ma, Phys. Lett. {\bf B865}, 139474 (2025).
\bibitem{m14} E. Ma, Phys. Rev. Lett. {\bf 112}, 091801 (2014).
\bibitem{fn79} C. D. Froggatt and H. B. Nielsen, Nucl. Phys. {\bf B147}, 277 
(1979).
\bibitem{fm14} S. Fraser and E. Ma, Europhys. Lett. {\bf 108}, 11002 (2014).
\bibitem{CMS25} A. Hayrapetyan {\it et al.} (CMS Collaboration), Phy. Rev. 
{\bf D112}, 032008 (2025). 
\bibitem{gjk21} P. Gutierrez, R. Jain, and C. Kao, Phys. Rev. {\bf D103}, 
115020 (2021).
 
\end{thebibliography}

\end{document}